\documentclass[aps, prd, twocolumn, showpacs, superscriptaddress, groupedaddress]{revtex4} 
\usepackage{tabularx}
\usepackage{array}
\usepackage{mathtools}
\usepackage{dcolumn}
\usepackage{color}
\usepackage{amsmath}
\usepackage{newtxtext}
\usepackage{graphicx}	
\usepackage{amssymb}
\usepackage{dcolumn}
\usepackage{ulem}
\usepackage{color}
\usepackage{leftidx}
\usepackage{subfigure, rotating, bm, array}
\usepackage[pagebackref=false, colorlinks=true]{hyperref}
\hypersetup{
linkcolor=blue,     
citecolor=blue,     
urlcolor=blue}      
\begin{document}
\title{Future-null singularity due to gravitational collapse}
\author{Ashok B. Joshi}
\email{gen.rel.joshi@gmail.com}
\affiliation{PDPIAS,
Charotar University of Science and Technology, Anand- 388421 (Guj), India.}
\author{Karim Mosani}
\email{kmosani2014@gmail.com}
\affiliation{Department of Mathematics, University of Tübingen, Auf der Morgenstelle 10, 72076 Tübingen, Germany.}
\author{Pankaj S. Joshi}
\email{psjcosmos@gmail.com}
\affiliation{International Centre for Space and Cosmology, School of Arts and Sciences, Ahmedabad University, Ahmedabad-380009 (Guj), India.}
\date{\today}

\begin{abstract}
In the case of unhindered gravitational collapse of matter cloud governed by the Lemaitre-Tolman-Bondi (LTB) spacetime, the end-state singularity is either locally visible, globally visible or completely hidden. We have a past-null singularity in the first two cases while a future-spacelike singularity in the last case. Here, we show an example of a gravitational collapse model whose end-state is a future-null-singularity (it has a causal property that is unlike the cases involving LTB spacetime).  We depict such a distinct causal structure of the singularity by conformal diagrams. 
\bigskip

$\boldsymbol{key words}$: Singularity, Causal structure of spacetime, Gravitational collapse
\end{abstract}
\maketitle

\section{Introduction}
The contraction of an astronomical body under its own gravitational influence is called gravitational collapse. When the mass of the star is above a certain limit, one obtains an unhindered gravitational collapse that gives rise to a singularity. In addition to the existence of incomplete causal curves, we identify the singularity by the blowing up of divergence of curvature-invariant quantities 
\cite{wald}. In 1939, Oppenheimer and Snyder studied the gravitational collapse of a spherically symmetric spatially homogeneous matter field with zero pressure
\cite{OppenheimerSnyder39}. 
Such collapse ends in a singularity covered by an event horizon.
One could argue that such a singularity is merely an artefact of spherical symmetry and that dropping the assumption of spherical symmetry could resolve the singularity. However, Penrose and Hawking showed the occurrence of singularities (identified by the existence of incomplete causal curves) under generic conditions. Penrose proposed what is now known as the cosmic censorship conjecture
\cite{Penrose:1969pc}. 
The weak version of cosmic censorship states: A spacetime singularity can never be visible to asymptotic observers. In other words, it cannot be globally visible. The strong version states: All physically reasonable spacetimes are globally hyperbolic, i.e., a singularity cannot even be locally naked. Here, a globally naked singularity is identified by the existence of past-incomplete causal geodesic that is future-complete, while a locally naked singularity is identified by the existence of past-incomplete causal geodesic that is also future-incomplete.

Let $\mathcal{M}$ be a spacetime with Lorentzian signature $+2$, and let $\mathcal{O}\subset \mathcal{M}$ be open. A \textit{congruence} $\mathcal{O}$ is a family of curves such that $\forall$ $p\in \mathcal{O}$, $\exists$ precisely one curve in this family that intersects $p$. The tangents to a congruence yield a vector field in $\mathcal{O}$. Conversely, every continuous vector field generates a congruence of curves. $\mathcal{O}$ is smooth if the corresponding vector field is smooth \cite{wald,Hawking,Joshi:1987wg}.

Consider a smooth congruence of null geodesics yielding a vector field $k:\mathcal{M}\to T\mathcal{M}$. The trace of the null Weingarten map 
$$\mathcal{W}: V_p~\to V_p~: ~\Bar{X}\mapsto \mathcal{W}(\Bar{X}):=\nabla_{\Bar{X}}k:=\overline{\nabla_{X}k},$$
is called the expansion of a null geodesic congruence at a point $p\in \mathcal{O}$.
Here $V_p$ is the two-dimensional fibre of the quotient set $T\mathcal{H}/\sim$ of the tangent bundle 
$$T\mathcal{H}:=\{Z\in T\mathcal{O}~\vert~g~(Z,k)=0\},$$ 
with the equivalence relation $\sim$ defined as follows: 
$$
\forall ~X, Y\in T\mathcal{O}, ~X\sim Y \iff X-Y\propto k.$$
$\Bar{X}\in V_p$ denotes the equivalence class of $X\in T\mathcal{H}$ for the above equivalence relation. A \textit{trapped surface} $T\subset H$ is then defined in terms of congruence as a compact two-dimensional smooth spacelike submanifold with the property that the expansion of both the families of future-directed null geodesics orthogonal to $T$  is $<0$ everywhere. A \textit{marginally trapped surface} is then defined in terms of the trapped surface with a relaxation that now the expansion of both the families of future-directed null geodesics orthogonal to $T$ is $\leq 0$ $\forall$ points in $T$, rather than strictly negative. A \textit{marginally outer trapped surface} is then defined by relaxing further the definition of marginally trapped surface such that the expansion of only ``outgoing" future-directed null geodesics orthogonal to $T$ is $\leq 0$ $\forall$ points in $T$ (Outgoing family: Family of null geodesics orthogonal to $T$ satisfying $g(k,N)\geq 0$, where $k$: Normal to null geodesics, $N$: Normal to $T$ in $S$ (Spacelike hypersurface containing $T$) that points outwards from $T$). A \textit{trapped region} is defined as a closed subset $C$ of partial Cauchy surface $S$ ($S\subset \mathcal{M}$: A spacelike hypersurface that intersects every causal curve $\gamma\subset \mathcal{M}$ at most once) that forms a three-dimensional manifold with a boundary such that its two-dimensional boundary $\partial C$ is a marginally outer trapped surface. Finally, the \textit{apparent horizon} is defined as the boundary $\partial C$ of the trapped region.

In the case of a dynamical spacetime that admits a singularity, e.g., gravitational collapse, the spacetime's causal property relates to the evolution of the apparent and the event horizon. As far as dust collapse is concerned, the introduction of spatial inhomogeneity in the density of the collapsing cloud influences the evolution of these horizons such that the end-state singularity becomes locally or globally naked
\cite{Eardley:1978tr, Dwivedi:1996wf, Mosani:2020mro}. 

\textcolor{black}{A spacetime singularity can be classified into five types based on the following two characteristics: 1) The causal structure of spacetime (globally hyperbolic or not) admitting the singularity, and 2) The existence of past/future complete/incomplete timelike/null geodesics \cite{Longair:1974fq}. The classification is as follows:
\begin{enumerate}
      \item \textit{Past spacelike singularity}: The spacetime is timelike as well as null geodesically past incomplete and globally hyperbolic; e.g. Schwarzschild white hole singularity.
      \item \textit{Future spacelike singularity}: The spacetime is timelike and null geodesically future incomplete and globally hyperbolic; e.g. Schwarzschild black hole singularity.
    .\item \textit{Past null singularity}: The spacetime is timelike as well as null geodesically past incomplete and not globally hyperbolic; e.g. locally/globally visible singularities in LTB spacetime.
    \item \textit{Future null singularity}:  The spacetime is timelike as well as null geodesically future incomplete and globally hyperbolic.
     \item \textit{Timelike singularity}: The spacetime is timelike geodesically complete (though not null), but $\exists$ at least one incomplete radial null geodesic or incomplete non-geodesic timelike curve \cite{Hawking}. Additionally, the spacetime is not globally hyperbolic.
\end{enumerate}
}

The authors in \cite{Ortiz:2011jw} consider a collapsing spherical dust cloud, generate the corresponding conformal diagram, and categorize the causal structure of the resulting singularity. The author in \cite{Kommemi:2011wh} has given the complete classification of future singularities in spacetimes that are solutions of the Einstein-Maxwell-Klein-Gordon equation.

Here, we show that a future null singularity can be obtained as an end-state of gravitational collapse of a physically reasonable matter cloud. To achieve this, we glue the static spacetime first introduced in \cite{Joshi:2020tlq} with the spatially homogeneous FLRW spacetime having non-zero pressure. The gist of  \cite{Joshi:2020tlq} is as follows: The general consensus is that in the case of static spherically symmetric spacetime admitting a singularity, its shadow is formed due to the presence of a photon sphere. In the abovementioned article, the authors give an example of the spacetime that admits a singularity but not a photon sphere; however, the singularity has a shadow.
%

The structure of the paper is as follows. In section (\ref{sec2}), we rederive the equations governing the dynamics of the apparent horizon, the singularity curve, and the event horizon curve for the gravitational collapse governed by LTB spacetime with the exterior Schwarzschild metric. This reviews the well-known formation of future spacelike singularity and past null singularity due to gravitational collapse in spatially homogeneous and inhomogeneous Lemaitre-Tolman-Bondi spacetimes, respectively. In section (\ref{sec3}), we use the Israel junction condition
\cite{Israel:1966rt} 
to glue the static spacetime first introduced in \cite{Joshi:2020tlq} (that we call the exterior spacetime) with the spatially homogeneous FLRW spacetime having non-zero pressure (that we call the interior spacetime). We show that the singularity in the static spacetime of \cite{Joshi:2020tlq} is future null (using a conformal diagram). The union of two spacetimes hence models a dynamical spacetime which ends in a future-null singularity. This glueing proves that just like the formation of past-null and future-spacelike singularities, future-null singularities can also be obtained in gravitational collapse. In other words, a static spacetime admitting future-null singularity can be obtained from dynamical spacetime (that does not admit a timelike Killing vector field), making it physically more relevant since all the structures in the universe are obtained from dynamical evolution. Finally, we end the paper with concluding remarks. Throughout the paper, we take $G=c=1$.

\section{Lemaitre-Tolman-Bondi spacetime}\label{sec2}
Consider the gravitational collapse of a spherically symmetric inhomogeneous perfect fluid.  The components of the stress-energy tensor in the coordinate basis $\{dx^{\mu} \bigotimes \partial_{\nu} \vert 0 \leq \mu, \nu \leq 3\}$ of the comoving coordinates $(t,r,\theta,\phi)$  are given by
\begin{equation}
    T^{\mu}_{\nu}=\textrm{diag}\left(-\rho,p,p,p\right),
\end{equation}
where $\rho=\rho(t,r)$ and $p=p(t,r)$ are the density and the isotropic pressure of the collapsing matter cloud, respectively. The corresponding spacetime metric is written as follows:
\begin{equation}
ds^2 = -dt^2 + R^{\prime 2}dr^2 + R^2 d\Omega ^2 \label{spacetime}
\end{equation}
where $d\Omega^2$ is the line element of the two-sphere, $R=R(t,r)$ and the superscript $'$ denotes the partial derivative of the function with respect to radial coordinate $r$. For such spacetime, we have
\begin{equation}
\rho = \frac{F^{'}}{ R^{'} R^2}\label{denspres}
\end{equation}
and
\begin{equation}\label{efep}
     p= - \frac{\dot{F}}{\dot{R} R^2},
\end{equation}
where
\begin{equation}
F= \dot{R}^2 R. \label{Epf}
\end{equation}
The superscript dot denotes the partial derivative of the function with respect to time coordinate $t$. $F=F(t,r)$ is called the Misner-Sharp mass function. A collapsing spherical solid ball contained in the initial data is made of concentric spherical shells, each of which is identified by a radial coordinate $r$. In the case of dust, we have $p=0$. Eq. (\ref{efep}) then implies that $F=F(r)$. In such scenario, one can integrate Eq. (\ref{Epf}) to obtain
\begin{equation}
R(t,r)  = \left( r^\frac{3}{2} - \frac{3}{2} \sqrt{F(r)}t\right)^\frac{2}{3}. \label{physicalr}
\end{equation}
We define the \textit{scaling function} $a(t,r)$ as the ratio
\begin{equation}
    a(t,r)=\frac{R(t,r)}{r}, \label{scalef}
\end{equation}
and rewrite Eq. (\ref{physicalr}) to obtain the \textit{time curve}
\begin{equation}\label{timecurve}
    t(r,a)=\frac{2r^{3/2}}{3\sqrt{F(r)}}\left(1-a^{3/2}\right).
\end{equation}
Given a spherical shell of fixed radial coordinates, as one evolves the initial data, the physical radius of this shell decreases and becomes zero. The corresponding comoving time $t_s(r)$ is obtained by substituting $R(t_s,r)=0$ in Eq. (\ref{physicalr}) or $a=0$ in Eq. (\ref{timecurve}) as
\begin{equation}
t_{s}(r) =  \frac{2r^{\frac{3}{2}}}{3\sqrt{F(r)}}. 
\end{equation}
We call this function the \textit{singularity curve}. As mentioned in the introduction, the expansion of the outgoing null geodesic congruence vanishes on the apparent horizon. In terms of the metric components (refer Eq. (\ref{spacetime})), it is 
\begin{equation}
    \theta= \frac{2}{R}\left(1- \sqrt{\frac{F}{R}}\right).
\end{equation}
$\theta=0$ is equivalent to $F=R$ from the above equation. Hence, the apparent horizon for such spacetime is the set
\begin{equation}
    \mathcal{A}=\left\{ a\in S\hspace{5pt} : \hspace{5pt} F\vert_a=R\vert_a \right\},
\end{equation}
where $S$ is a partial Cauchy surface (We choose $S$ to be a constant $t$ time slice. In other words, $S$ should be such that $\partial_t$ is the (global) unit normal vector field).
Geometrically, $\mathcal{A}$ is a two-dimensional sphere satisfying 
\begin{equation}
H+{^{\mathcal{A}}}\textrm{tr} K=0.
\end{equation}
Here, $H$ is the mean curvature of $\mathcal{A}$ in $\left(S, {^S} g\right)$, $K$ is the extrinsic curvature of $S$ in $\mathcal{M}$, ${^{\mathcal{A}}}\textrm{tr} K$ is the trace of $K$ over $\mathcal{A}$, or in other words, the trace of $K\vert_{T\mathcal{A}\times T\mathcal{A}}$ with respect to the induced metric ${^{\mathcal{A}}}g$ on $\mathcal{A}$. 

\begin{figure}
\centering
{\includegraphics[width=62mm]{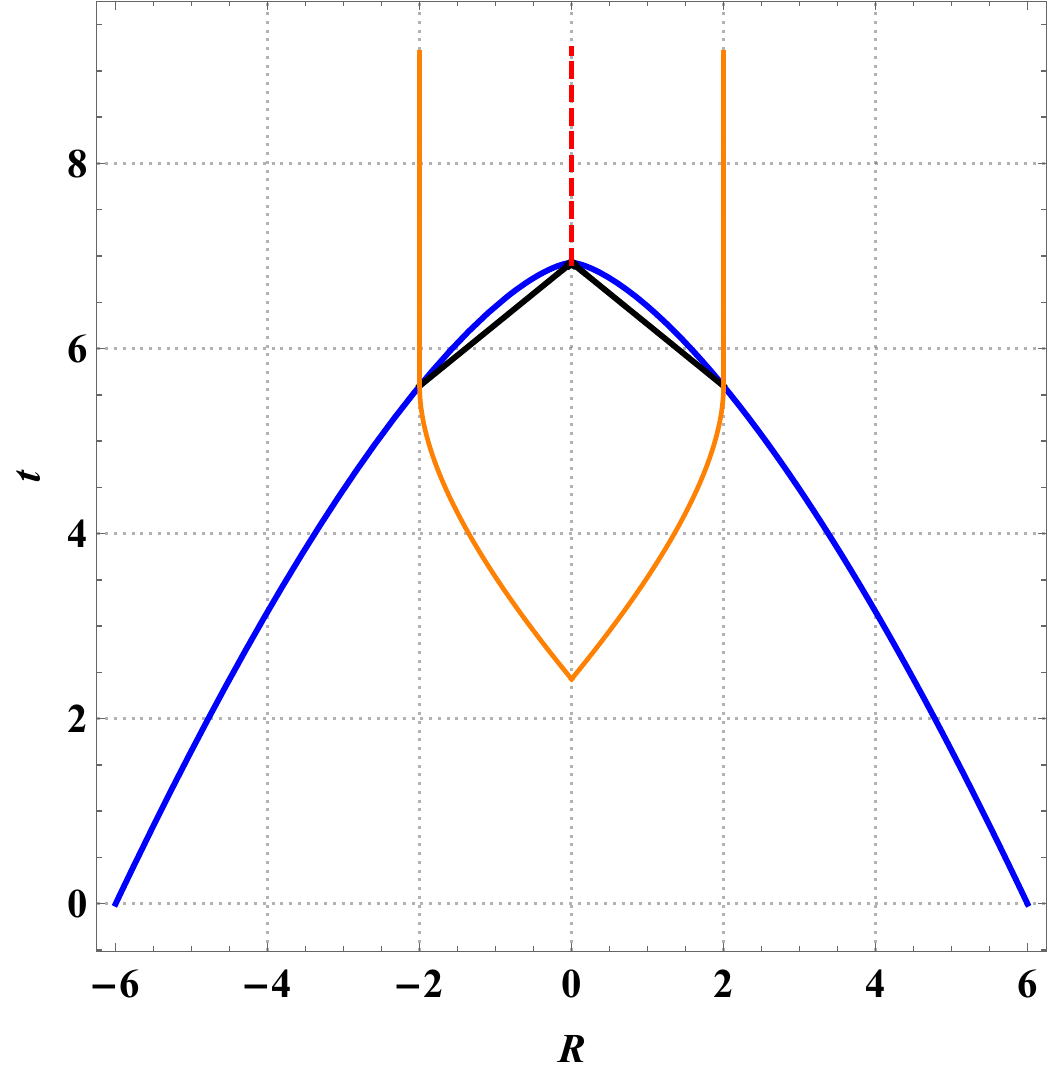}}
 \caption{The gravitational collapse of spatially-homogeneous dust governed by the LTB spacetime. The solid yellow, blue, and black curves represent the event horizon, the boundary of the collapsing cloud, and the apparent horizon, respectively.}\label{HomoLTB}
\end{figure}
In the case of spherical symmetry, the apparent horizon is a spherical ball embedded in $S$ whose points have a fixed radial coordinate. As one evolves the initial data $S$, $\mathcal{A}$ evolves (the radial coordinate of points in $\mathcal{A}$ evolves). This evolution is obtained from (\ref{physicalr}) along with equating $F=R$ to get 
\begin{equation}\label{ltbtimeah}
    t_{ah}(r) =\frac{2}{3}\left(\frac{r^{3/2}}{\sqrt{F}}-F\right).
\end{equation}
We call this function the \textit{apparent horizon curve}. It gives us the relation between the radial coordinate of points in $\mathcal{A}$ and the comoving time $t$.

The collapsing perfect fluid spacetime Eq. (\ref{spacetime}) can be
matched smoothly with the exterior Schwarzschild spacetime so that the union validly solves Einstein’s field equation. By smooth matching, we mean the satisfaction of the Darmios-Israel junction conditions \cite{Israel:1966rt}. 
It states that at the hypersurface (in our case: a timelike hypersurface identified by $r=r_c$, where $r_c\in \mathbb{R}^+$), the induced LTB metric and the induced Schwarzschild metric should be the same. Secondly, the extrinsic curvature of this hypersurface in LTB spacetime and in Schwarzschild spacetime should be the same.

In the case of static spacetime, the event horizon is also the apparent horizon. Hence, one can obtain the evolution of the event horizon by solving the differential equation
\begin{equation}
    \frac{dt}{dr} = R^{'}(t,r), \label{nullgeodesicsltb}
\end{equation}
with the initial condition that at the boundary of the collapsing cloud, it should coincide with the apparent horizon, i.e. $F\vert_a=R\vert_a$, where $a$ has radial coordinate $r=r_c$. The solution to this initial value problem gives us the \textit{event horizon curve} and is denoted by $t_{eh}(r)$.

Regarding the formation of at least locally visible singularity, we have the following statement: Consider an unhindered gravitational collapse of a spherically symmetric perfect fluid. The singularity formed as an end state of such collapse is at least locally naked if and only if $\exists$ $X_0\in \mathbb{R}^{+}$ as a root of $V(X)$, where 
\begin{widetext}
    \begin{equation}\label{rootequation}
    V(X)=X-\frac{1}{\alpha}\left(X+\sqrt{\frac{F_{0}(0)}{X}}\left(\chi_1(0)+2r\chi_2(0)+3r^2\chi_3(0)\right)r^{\frac{5-3\alpha}{2}}\right)\left(1-\sqrt{\frac{F_0(0)}{X}}r^{\frac{3-\alpha}{2}}\right).
    \end{equation}
   Here
    \begin{equation}
        \alpha \in \left\{\frac{2n}{3}+1;\hspace{0.2cm} \hspace{0.2cm} n\in \mathbb{N} \right\}, \hspace{0.5cm}
    \chi_i(v)=\frac{1}{i!}\frac{\partial ^i t(r,v)}{\partial r^i}\bigg |_{r=0}, \hspace{0.5cm} F(r,v)=\sum_{i=0}^{\infty}F_i(v)r^{i+3},
    \end{equation}
\end{widetext}
and $t=t(r,v)$ is the time curve \cite{Joshi:1993zg}. 

\begin{figure*}
\centering
\subfigure[]
{\includegraphics[width=62mm]{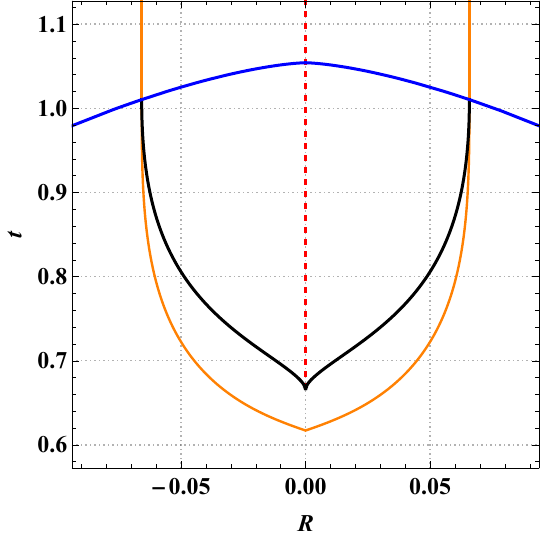}\label{ltbcollapseInHomo3}}
\hspace{1cm}
\subfigure[]
{\includegraphics[width=62mm]{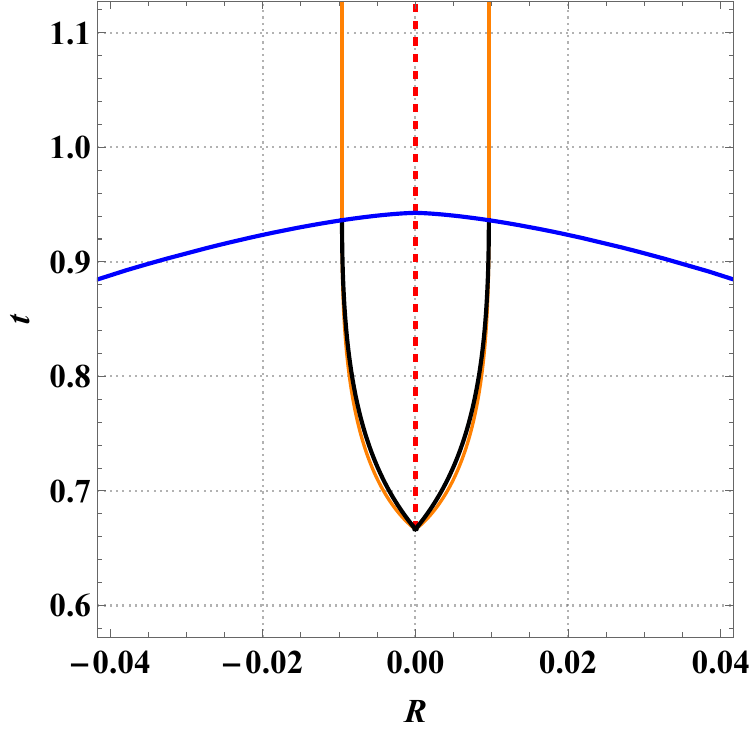}\label{ltbcollapseInHomo5}}
 \caption{The gravitational collapse of spatially-inhomogeneous dust governed by the LTB spacetime. The solid yellow, blue, and black curves represent the event horizon, the boundary of the collapsing cloud, and the apparent horizon, respectively: (a) $F(r)=F_0 r^3+F_3 r^5$. The singularity is locally visible \textcolor{black}{($F_{0} =1$ and $F_{2} = -2$)}, (b) $F(r)=F_0 r^3+F_3 r^6$. The singularity is globally visible \textcolor{black}{($F_{0} =1$ and $F_{3} = -26$)}.
 }\label{InHomo}
\end{figure*}


To depict examples of past null singularities, we consider two different Misner-Sharp mass functions $F_a$ and $F_b$ given by
    \begin{equation}\label{exampleltbg}
        F_a(r)=F_0 r^3+ F_3 r^6; \hspace{0.3cm} F_0>0, \hspace{0.3cm} F_3<0. 
    \end{equation} 
and    
    \begin{equation}\label{exampleltbl}
        F_b(r)=F_0 r^3+ \textcolor{black}{F_2 r^5}; \hspace{0.3cm} F_0>0, \hspace{0.3cm} F_2<0. 
    \end{equation}
    respectively.
    Corresponding $V(X)$ (Eq. (\ref{rootequation})) for such mass functions are obtained as: 
    \begin{equation}\label{rooteqnltb}
    \begin{split}
         V_a(X)=  2 X^2 + \sqrt{F_0}X^{3/2} - 3\sqrt{F_0} \chi_3(0) \sqrt{X} \\ + 3F_0 \chi_3(0),
    \end{split}
    \end{equation}
    where 
    \begin{equation}
        \chi_3(0)= -\frac{1}{2} \int_{0}^{1} \frac{F_3/a}{\left(F_0/a \right)^{3/2}}~ da,
    \end{equation}
    and
    \begin{equation}\label{rooteqnltb2}
    \begin{split}
         V_b(X)= 4 X^2 - 6\sqrt{F_0} \chi_2(0) \sqrt{X}
    \end{split}
    \end{equation}
   where
    \begin{equation}
        \chi_2(0)= -\frac{1}{2} \int_{0}^{1} \frac{F_2/a}{\left(F_0/a \right)^{3/2}}~ da.
    \end{equation}
Setting $F_0=1$, Eq. (\ref{rooteqnltb}) has positive real roots if and only if $F_3<\sim-25.967$, in which case, the singularity is visible. Similarly, Eq. (\ref{rooteqnltb2}) has a positive real root if and only if $F_2<0$, in which case the singularity is visible.

For the singularity to be globally visible, apart from the above-mentioned criteria, the following equality should be satisfied:
\begin{equation}\label{ehs}
    t_{eh}(0)=t_s(0).
\end{equation}
It can be seen numerically that in the former example (Eq. (\ref{exampleltbg})), Eq. (\ref{ehs}) is always satisfied if Eq. (\ref{rooteqnltb}) has a positive real root. Hence, the visible singularity is globally visible. In the latter example, if Eq. (\ref{rooteqnltb2}) has a positive real root, then the singularity is either locally visible or globally visible depending on the choice of the coefficients $F_0$ and $F_2$. 


Figs.~(\ref{HomoLTB}) and (\ref{InHomo}) depict the spacetime diagrams consisting of the evolution of the apparent horizon and the event horizon in the case for which the LTB collapse leads to a locally hidden singularity, a locally visible singularity, and a globally visible singularity, respectively. We now proceed to depict the formation of future-null singularity as an end-state of gravitational collapse. 

\section{Future-null singularity admitting spacetime}\label{sec3}
Consider the gravitational collapse of a spherically symmetric spatially homogeneous perfect fluid.  The components of the stress-energy tensor in the coordinate basis $\{dx^{\mu} \bigotimes \partial_{\nu} \vert 0 \leq \mu, \nu \leq 3\}$ of the comoving coordinates $(t,r,\theta,\phi)$  are given by
\begin{equation}
    T^{\mu}_{\nu}=\textrm{diag}\left(-\rho,p,p,p\right),
\end{equation}
where $\rho=\rho(t)$ and $p=p(t)$ are the spatially homogeneous density and the isotropic pressure of the collapsing matter cloud. 
The corresponding spacetime metric is written just as Eq. (\ref{spacetime}). In the case of dust, the Misner-Sharp mass function is conserved inside a shell of radial coordinate $r$, and the exterior metric is the Schwarzschild spacetime. Hence, for the exterior spacetime to be non-vacuum, we choose the Misner-Sharp mass function such that it is not conserved inside a shell of the fixed radial coordinate. Hence, it is a function of both $t$ and $r$ coordinates. The scaling function $a$ is a function of only $t$ coordinate in the case of spatially homogeneous collapse.



\begin{figure}
\centering
{\includegraphics[width=70mm]{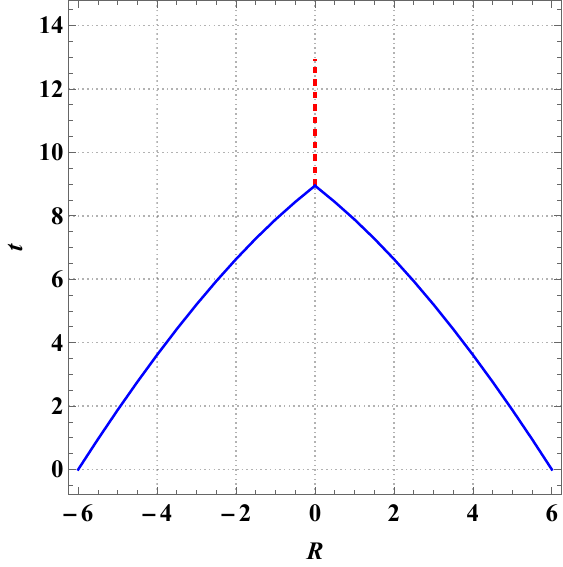}\label{ltblikecollapse}}
 \caption{The gravitational collapse of spatially-homogeneous perfect fluid governed by the FLRW spacetime glued to an asymptotically flat non-vacuum spacetime (Eq. (\ref{schajst})). The solid blue curve represents the boundary of the collapsing cloud. The event horizon and the apparent horizon are absent.
 }\label{fig3}
\end{figure}


\begin{figure*}
\centering
\subfigure[]
{\includegraphics[width=53mm]{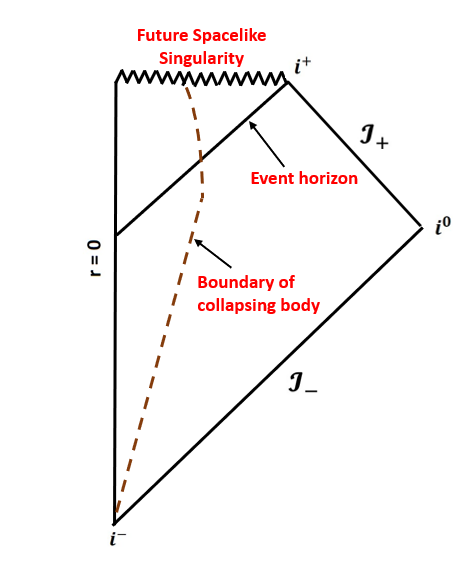}\label{ltbpenrose}}
\hspace{2.5cm}
\subfigure[]
{\includegraphics[width=60mm]{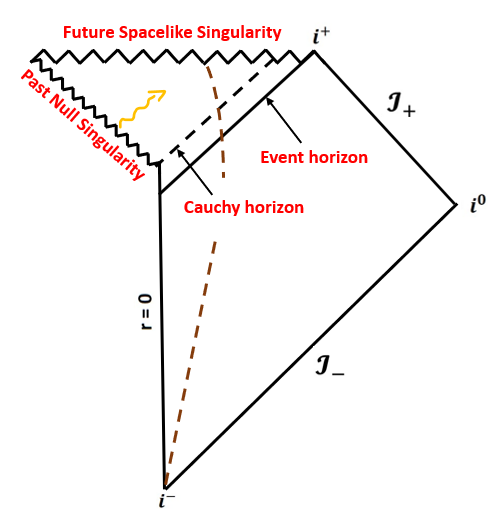}\label{nullpenrose2}}\\
\subfigure[]
{\includegraphics[width=60mm]{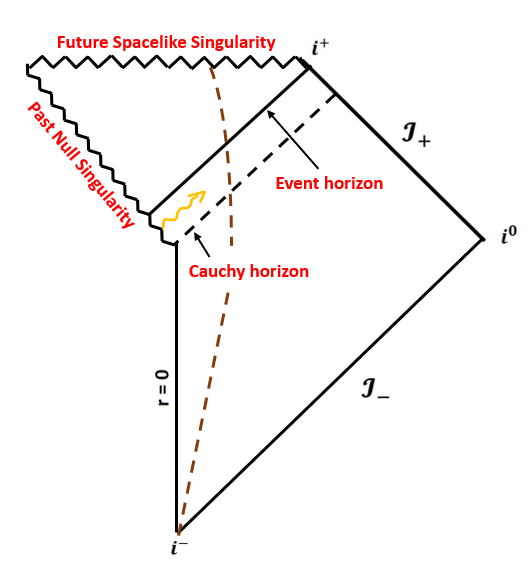}\label{nullpenrose1}}
\hspace{2.5cm}
\subfigure[]
{\includegraphics[width=56mm]{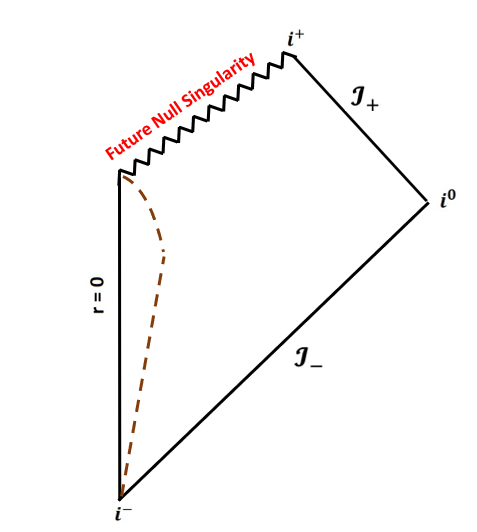}\label{nullpenrose}}
 \caption{Conformal diagram of four different causal structures of the singularities formed due to gravitational collapse (depicted by brown dashed curve). (a) Schwarzschild singularity formed due to spatially homogeneous gravitationally collapsing dust glued to exterior Schwarzschild spacetime, (b) Locally naked singularity formed due to spatially inhomogeneous gravitationally collapsing dust glued to exterior Schwarzschild spacetime, (c) Globally naked singularity formed due to spatially inhomogeneous gravitationally collapsing dust glued to exterior Schwarzschild spacetime, (d) Future null singularity formed due to spatially homogeneous gravitationally collapsing perfect fluid glued to exterior asymptotically-flat non-vacuum spacetime (\ref{schajst}).
 }\label{penrosed}
\end{figure*}
Unlike the collapsing dust LTB spacetime, the collapsing non-zero pressured FLRW spacetime is matched with the exterior spherically symmetric asymptotically flat non-vacuum spacetime discussed in 
\cite{Joshi:2020tlq}, instead of Schwarzschild spacetime.
This exterior spacetime metric is given in Schwarzschild coordinates by
\begin{equation}\label{schajst}
    ds^2=-\frac{dt^2}{\left(1+\frac{M}{r}\right)^2}+\left(1+\frac{M}{r}\right)^2 dr^2+r^2d\Omega^2,
\end{equation}
where $M$ is a positive constant. 

In Eddington–Finkelstein coordinates, it is expressed as
\begin{equation}
    ds^2 = -\left(1 - \frac{2D( \mathcal{R})}{\mathcal{R}}\right) d\nu^2 - 2 d\nu d\mathcal{R}+ \mathcal{R}^2 d\Omega^2 \label{nullmatch}
\end{equation}
where,
\begin{equation}
    D(\mathcal{R}) = \frac{M \mathcal{R}(M + 2\mathcal{R})}{2 (M + \mathcal{R})^{2}} \label{nullmass}
\end{equation}
Here $\mathcal{R}$ is the radial coordinate of the exterior spacetime, and $\nu$ is retarded null coordinate. In the coordinate basis $\{dx^{\mu} \bigotimes \partial_{\nu} \vert 0 \leq \mu, \nu \leq 3\}$ of the Schwarzschild coordinates, the stress-energy tensor corresponding to Eq. (\ref{nullmatch}) is given by
\begin{equation}
T=\textrm{diag}\{\epsilon, -\epsilon, \mathcal{P}, \mathcal{P}\}
\end{equation}
where $\epsilon>0$. 
In the orthonormal basis $\{e_{(i)}\bigotimes e_{(j)}\hspace{5pt} \vert \hspace{5pt} 0\leq i,j \leq 3\}$, where 
\begin{equation}
    e_{(i)}=\frac{\partial_{i}}{\sqrt{g_{ii}}},
\end{equation}
it is
\begin{equation}
T=\textrm{diag}\{\epsilon, -\epsilon, \mathcal{P}, \mathcal{P}\},
\end{equation}
(Here $g_{ii}$ is the $ii$'th component of the metric tensor in the coordinate basis $dx^i\bigotimes dx^j$ of the Schwarzschild coordinates).
The stress-energy tensor of generalized Vaidya spacetime in the abovementioned orthonormal basis is written as
\begin{equation}
T=
    \begin{pmatrix}
\frac{\Bar{\epsilon}}{2}+\epsilon & \frac{\Bar{\epsilon}}{2} & 0 & 0\\
\frac{\Bar{\epsilon}}{2} & \frac{\Bar{\epsilon}}{2}-\epsilon & 0 & 0 \\
0 & 0 & \mathcal{P} & 0\\
0 & 0 & 0 & \mathcal{P}.
\end{pmatrix}
\end{equation}
Hence, the exterior spacetime with metric Eq. (\ref{schajst}) is a special case of generalized Vaidya spacetime with $\Bar{\epsilon}=0$ \cite{Wang:1998qx}.

Now, we briefly discuss the Israel Junction condition we employ to match the interior and exterior spacetime: The collapsing perfect fluid spacetime Eq. (\ref{spacetime}) (that we call here as ($\mathcal{M}_1, g_1$)) can be matched ``smoothly" with the exterior Schwarzschild spacetime (that we call here as ($\mathcal{M}_2,g_2$)) at the timelike hypersurface $\Sigma$ (identified by the collapsing shell of largest comoving radius) embedded in both $\mathcal{M}_1$ and $\mathcal{M}_2$.
The new spacetime is then a valid solution of Einstein's field equations. By ``smooth" matching, we mean the following:  The oriented boundaries $\partial \mathcal{M}_1$ and $\partial \mathcal{M}_2$ respectively should be such that $\partial \mathcal{M}_1$ is diffeomorphic to $\partial \mathcal{M}_2$, i.e. $\partial \mathcal{M}_1\cong \partial \mathcal{M}_2\cong \Sigma$. A new spacetime $(\mathcal{M},g)$ is then constructed such that $\mathcal{M}$ is a disjoint union of $\mathcal{M}_1$ and $\mathcal{M}_2$. $\partial \mathcal{M}_1$ and $\partial \mathcal{M}_2$ are embedded in $\mathcal{M}_1$ and $\mathcal{M}_2$, respectively, and their points are identified such that certain conditions (that we call Israel's conditions) are satisfied
\cite{Israel:1966rt, Mena_11}. Let $\leftidx{^{\Sigma}}{g}{_1}$ denote the metric induced on $\Sigma \hookrightarrow \mathcal{M}_1$ and $\leftidx{^{\Sigma}}{g}{_2}$ denote the metric induced on $\Sigma \hookrightarrow \mathcal{M}_2$. Let
$$
K_1: T\Sigma \times T\Sigma \to ~C^{\infty}\left(\Sigma\right):~(X,Y)\mapsto g_1\left(\nabla_X\nu, Y\right),
$$
and
$$
K_2: T\Sigma \times T\Sigma \to ~C^{\infty}\left(\Sigma\right):~(X,Y)\mapsto g_2\left(\nabla_X\nu, Y\right),
$$
be the extrinsic curvatures of $\Sigma \hookrightarrow \mathcal{M}_1$ and $\Sigma \hookrightarrow \mathcal{M}_2$ respectively (here $\nu\in T\mathcal{M}$ is the global unit normal vector field of $\Sigma$). The Israel junction conditions are as follows: 1) $\leftidx{^\Sigma}{g}{_1}\equiv\leftidx{^\Sigma}{g}{_2}$, and 2) $K_1\equiv K_2$.

\textcolor{black}{Matching these first and second fundamental forms for the interior and exterior metric on the matching surface $\Sigma$ identified by the radial coordinate $r=r_c$ gives the following four equations \cite{Goswami:2007na}:
\begin{equation}
    \mathcal{R} = R(t, r_{c}) = r_{c}  a(t), \label{matchedr}
\end{equation}}

\begin{equation}
    \left(\frac{d\nu}{dt}\right)_{\Sigma} = \frac{1 +  \dot{\mathcal{R}}}{\left(1- \frac{F(t,r_{c})}{\mathcal{R}}\right)}, \label{matchnullco}
\end{equation}

\begin{equation}
     F(t,r_{c}) = 2 D(\mathcal{R}),\label{massequal}
\end{equation}
\hspace{0.5cm}
and
\begin{equation}
   D(\mathcal{R})_{,\mathcal{R}} = \frac{ F(t,r_{c})}{2\mathcal{R}} + \mathcal{R} \ddot{\mathcal{R}}. \label{match2}
\end{equation}
Eqs. (\ref{matchedr}) and (\ref{matchnullco}) are obtained from matching the first fundamental forms, while Eqs.~(\ref{massequal}) and (\ref{match2}) are obtained from matching the second fundamental forms at $\Sigma$
\cite{psjoshi2}.
From Eqs.~(\ref{Epf}), (\ref{nullmass}), (\ref{matchedr}) and (\ref{massequal}), and the fact that $\dot a<0$,we obtain  
\begin{equation}
    \dot{a} = -\frac{\sqrt{M \left(M + 2 r_{c} a\right)}}{r_{c} \left(M +  r_{c} a\right)}. \label{soln}
\end{equation}
Solving this with the initial condition $a(t=0)=1$, and the constrain $\dot a<0$ $\forall$ $t\in [0,t_s)$  gives us 
\begin{equation}
    a(t) = \frac{1}{2 r_{c}}\left(-3M + \frac{M^{2}}{\psi(t)^{\frac{1}{3}}} + \psi(t)^{\frac{1}{3}}\right), \label{scalefector}
\end{equation}
where  
\begin{widetext}
\begin{align*}
    \psi(t) &= 9 M^3 + 24 M^2 r_{c} + 4 r_{c}^3 - 12 r_{c} \sqrt{M (M + 2 r_{c})} t - 24 \sqrt{M^3 (M + 2 r_{c})} t + 18 M (r_{c}^2 + t^2) +\\ 
    & \sqrt{-M^6 + (M^3 + 2 (2 M + r_{c})^2 (M + 2 r_{c}) - 
   12 (2 M + r_{c}) \sqrt{M (M + 2 r_{c})} t + 18 M t^2)^2}.
\end{align*}
\end{widetext}
Eqs.~(\ref{nullmass}), (\ref{massequal}) and (\ref{match2}) gives
\begin{equation}
    \ddot{\mathcal{R}} = - \frac{M \mathcal{R}}{(M + \mathcal{R})^{3}}. \label{matchaccelaration}
\end{equation}
which is again satisfied by $a(t)$ in Eq. (\ref{scalefector}).

We now have an example of the gravitational collapse (interior spatially-homogeneous perfect fluid spacetime (\ref{spacetime}) with the scaling function given by Eq. (\ref{scalefector}), matched smoothly with the exterior specific example of generalized Vaidya spacetime (\ref{nullmatch})) that gives rise to a future-null singularity. Such singularity is obtained at comoving time
\begin{equation}
    t_{s} = \frac{1}{3}\left((2M+r_{c})\sqrt{M+2r_{c}} -2M^{3/2}\right),
\end{equation}
that is achieved by substituting $a(t=t_s)=0$ in Eq. (\ref{scalefector}). For the scaling function given by Eq. (\ref{scalefector}), we can obtain the explicit expression of $F(t,r)$ using Eq. (\ref{Epf}). We can then see that $\nexists$ $(t,r)\in (0,t_s)\times (0,r_c)$ satisfying  $F=R$. This implies the absence of the apparent horizon and, hence, the event horizon.
Fig.~(\ref{fig3}) depicts the spacetime diagram of spatially homogeneous collapse giving rise to a future-null singularity. 
Fig.~(\ref{penrosed}) depicts the conformal diagram of four spacetimes with different causal structures (existence of future spacelike singularity, past null singularity: 1) locally visible, 2) globally visible, and finally, future null singularity).

\section{Concluding remark}\label{result}

In the collapsing spatially homogeneous LTB spacetime, the end-state is a future-spacelike singularity. In the collapsing spatially inhomogeneous LTB spacetime, the end-state is a past-null singularity. Here, we showed that the gravitational collapse can also lead to the formation of future-null singularity as an end-state. An interior collapsing FLRW spacetime with a time-dependent Misner-Sharp mass function ($F(t,r)=a~\dot{a}^2r^3$, where $a(t)$ is as shown in Eq. (\ref{scalefector})) glued to an exterior asymptotically flat non-vacuum spacetime first discussed in 
\cite{Joshi:2020tlq} gives rise to such singularities. 

\end{document}